\documentclass{iaus}
\usepackage{graphicx}
\usepackage{caption}

\title[] 
{Variability of Young Massive Stars in the Arches Cluster: Accurate Photometry with Adaptive Optics}

\author[]   
{K. Markakis$^1$$^,$$^2$
 \, A.Z. Bonanos$^1$$^,$$^2$
 \, G. Pietrzynski$^3$
 \, L. Macri$^4$
 \, K.Z. Stanek$^5$}

\affiliation{$^1$National Observatory of Athens, Institute of Astronomy \& Astrophysics, \\ I. Metaxa \& Vas. Pavlou St., P. Penteli 15236, Athens, Greece \\ {\tt markakis@astro.noa.gr}, {\tt bonanos@astro.noa.gr} \\[\affilskip]
$^2$K.M. \& A.Z.B. acknowledge support from the IAU and the European Commission \\for an FP7 Marie Curie International Reintegration Grant. \\[\affilskip]
$^3$Warsaw University Observatory, Al. Ujazdowskie 4, 00-478 Warszawa, Poland\\ Universidad de Concepci{\'o}n, Departamento de Astronomia,
Casilla 160-C, Concepci{\'o}n, Chile
\\[\affilskip]
$^4$Department of Physics \& Astronomy, Texas A\&M University, College Station, TX 77842-4242, USA\\[\affilskip]
$^5$The Ohio State University, 140 West 18th Avenue, Columbus, OH 43210, USA{}}

\pubyear{2011}
\volume{282}  
\pagerange{1--2}
\jname{From Interacting Binaries to Exoplanets:\\Essential Modelling Tools} 
\editors{A.C. Editor, B.D. Editor \& C.E. Editor, eds.}
\begin{document}

\maketitle

\begin{abstract}
We present preliminary results of the first near-infrared variability study of the Arches cluster, using adaptive optics data from NIRI/Gemini and NACO/VLT. The goal is to discover eclipsing binaries in this young (2.5 $\pm$ 0.5 Myr), dense, massive cluster for which we will determine accurate fundamental parameters with subsequent spectroscopy. Given that the Arches cluster contains more than 200 Wolf-Rayet and O-type stars, it provides a rare opportunity to determine parameters for some of the most massive stars in the Galaxy.
\end{abstract}

\firstsection 
\section{Introduction}
One of the most important questions is how massive can the most massive stars in the Universe be today. In other words what is the upper limit of the Initial Mass Function in the Universe. The Arches Cluster provides us with a unique opportunity to address this question. Being a young massive cluster which lies near the Galactic center, it is bound to contain massive eclipsing binary systems, which provide the means to accurately measure parameters of massive stars \cite [(Bonanos 2009)]{Bonanos09}. 
\vspace{-0.6cm}
\section{Datasets \& Reduction}
We used two datasets in the $K_s$ band. The first dataset was obtained with Gemini's NIRI infrared camera covering 8 nights from April to July of 2006. The NIRI data have undergone a linearity correction. The second dataset was obtained with the VLT's NACO infrared camera on 29 nights between June of 2008 and March of 2009. The reduction of the NIRI images was performed with the IRAF\footnote{IRAF is distributed by the NOAO, which are operated by the
Association of Universities for Research in Astronomy, Inc., under
cooperative agreement with the NSF.} Gemini v1.9 package while
the reduction of the NACO images was performed via the NACO reduction pipeline, based on ESO's Common Pipeline Library.
\firstsection
\section{Image Subtraction \& Photometry}
We tested four different methods in order to achieve accurate photometry. Initially, we began with the image subtraction package ISIS \cite[(Alard \& Lupton 1998]{Alardlupton98}\cite[, Alard 2000)]{Alard00}, which is optimal for detecting variables in crowded fields and IRAF's DAOPHOT \cite[(Stetson 1987)]{Stetson87} package. Both of these software packages use a mathematical PSF model in order to model the stellar light profile and produce large photometric errors (the order of 2-5 magnitude differences for the same object from frame to frame which is not physically acceptable) primarily because of the speckles that are being introduced by the use of adaptive optics. We concluded that it is impossible to fit a mathematical function on these speckles. In order to solve this problem we tried a different approach with the use of an empirical PSF. For this reason we used the StarFinder code \cite[(Diolaiti et al. 1999]{Diolaiti99}\cite[, 2000)]{Diolaiti00}. With the original version of the code we saw a big improvement as far as the photometric errors are concerned (the magnitude differences have dropped below 1 magnitude which is physically acceptable). However we were not able to identify any non-variable stars. The reason for this is that the StarFinder code does not allow for a spatially variable PSF option which is crucial in our case since the already imperfect correction by the adaptive optics degrades rapidly with increasing distance from the AO guide star. In order to improve our results further we used the version of the StarFinder code developed by Schoedel \cite[(2010)]{Schoedl10}. The main difference of this version is that it uses a local PSF by dividing the frame into several subframes with large overlap with each other. The PSF is considered to be stable across these subframes. Moreover the code performs photometry on each object with more than one PSF models (on most occasions the same object appears on more than one frames due to the large overlap) which helps the statistics of the actual counts value. Another interesting and helpful feature of Dr. Schoedel's approach is that the code performs photometry on a Wiener deconvolved version of the original image which favors the deblending of nearby sources in dense fields. After applying this method on our frames we saw further improvement on our results. The magnitude differences have now dropped even further (in the range of 0.2 to 0.6 magnitudes) but the problem of not finding stable stars remains. This behavior of our data may be explained by the underestimated errors produced by StarFinder (an issue that has been discussed by several researchers), by a possible contamination of our data (aborted nights, bad performance of the AO system etc.) or by a combination of the above.

\firstsection
\section{Future work \& Conclusions}
Currently we are trying to check our sample for possible contamination with bad quality frames and to better estimate the errors that the code produces. We conclude that the use of an empirical PSF is mandatory for accurate photometry on AO data. Moreover the use of Wiener deconvolution is very helpful when one works on crowded regions as it favors the deblending of nearby sources.


\end{document}